\documentclass[aps,showpacs,showkeys,prd,preprintnumbers,nofootinbib]{revtex4}
\usepackage{epsfig}
\usepackage{graphicx}
\usepackage{color}
\usepackage{bbm}                                                  

\definecolor{straw}{rgb}{1,1,0.50}

\newcommand{\nc}{\newcommand}
\nc{\beq}{\begin{equation}}  \nc{\eeq}{\end{equation}}
\nc{\bea}{\begin{eqnarray}}  \nc{\eea}{\end{eqnarray}}
\nc{\bpm}{\begin{pmatrix}}   \nc{\epm}{\end{pmatrix}}
\nc{\bit}{\begin{itemize}}   \nc{\eit}{\end{itemize}}
\nc{\bce}{\begin{center}}    \nc{\ece}{\end{center}}
\def\xx{{\bf x}}

\def\bcal{{\cal B}}

\def\ical{{\cal I}}
\def\lcal{{\cal L}}
\def\mcal{{\cal M}}

\def\ocal{{\cal O}}
\def\ucal{{\cal U}}
\def\zcal{{\cal Z}}
\def\gesim{\,{\raise-3pt\hbox{$\sim$}}\!\!\!\!\!{\raise2pt\hbox{$>$}}\,}
\def\lesim{\,{\raise-3pt\hbox{$\sim$}}\!\!\!\!\!{\raise2pt\hbox{$<$}}\,}
\def\vevof#1{\left\langle#1\right\rangle}
\def\half{\frac12}
\def\inv#1{\frac1{#1}}
\def\qwe{\delta}

\nc\ssm{{\rm SM}}
\nc\snp{{\rm NP}}
\nc{\du}{d_\ucal}
\nc{\dsm}{d_\ssm}
\nc{\bz}{{\bcal\zcal}}
\nc{\dbz}{{d_{\bz}}}
\nc{\lu}{\Lambda_\ucal}
\nc{\lun}{\Lambda_{\not\ucal}}
\nc{\ou}{\ocal_\ucal}
\nc{\osm}{\ocal_{SM}}
\nc{\obz}{\ocal_{\bz}}
\nc{\mpl}{M_{Pl}}
\nc{\tbz}{T_{\bz}}
\nc{\mc}{M_\ucal}
\nc{\cu}{C_\ucal}
\nc{\tf}{T_f}
\nc{\tfu}{T_{\ucal\hbox{\tiny-}f}}
\nc{\tfbz}{T_{\bz\hbox{\tiny-}f}}
\nc{\tsm}{T_\ssm}
\nc{\tnp}{T_\snp}
\nc{\tew}{T_{\rm EW}}
\nc{\tu}{T_\ucal}
\nc{\gsm}{g_\ssm}
\nc{\gu}{g_{\ucal}}
\nc{\gbz}{g_\bz}
\nc{\gir}{g_{\rm IR}}
\nc{\gnp}{g_\snp}
\nc{\gtot}{g_{\rm tot}}
\nc{\dg}{f}
\nc{\rsm}{\rho_{\rm SM}}
\nc{\ru}{\rho_\ucal}
\nc{\rbz}{\rho_\bz}
\nc{\rnp}{\rho_{\rm NP}}
\nc{\rtot}{\rho_{\rm tot}}
\nc{\co}{{\rm const.}}
\nc{\Pu}{P_\ucal}
\nc{\lm}{g}

\preprint{IFT-09-15 \cr UCRHEP-T482}

\begin{document}

\title{Uncosmology
}

\author{Bohdan GRZADKOWSKI}
\email{bohdan.grzadkowski@fuw.edu.pl}
\affiliation{Institute of Theoretical Physics,  University of Warsaw,
Ho\.za 69, PL-00-681 Warsaw, Poland}

\author{Jos\'e WUDKA}
\email{jose.wudka@ucr.edu}
\affiliation{Department of Physics and Astronomy, University of California,
Riverside CA 92521-0413, USA\\
and\\
Departamento de F\'\i sica Te\'orica y del Cosmos\\
Universidad de Granada E-18071, Granada, Spain}

\pacs{11.15.-q, 98.80.Cq}

\begin{abstract}
We discuss some cosmological features of a hypothetical
type of new physics characterized by begin asymptotically free
in the UV regime and conformally invariant in the IR. We show
that nucleosynthesis data generates non-trivial
constrains this type of models.
\end{abstract}

\maketitle

\section{The basic idea}

The ``unparticle'' 
proposal~\cite{Georgi:2007ek} is based on the assumption that
there is a type of new physics (NP) with the peculiar properties
of being asymptotically free in the ultraviolet (UV) and conformally 
invariant in the infrared (IR).
It is assumed that the NP  interacts
weakly with the standard model (SM)  through the exchange of a
set of heavy mediators of mass $ \mc $. Though not explicitly stated it
is also tacitly assumed that the theory will have
an ultraviolet  completion, whose details (aside form the
existence of the above-mentioned mediators) are left
unspecified.
The NP sector has two relevant high-energy scales:
$ \mc$ and $ \lu $ the scale
at which conformal invariance sets in. One must have
$ \mc > \lu $ and we will also assume $ \lu > v =246$GeV,
the electroweak scale.
The basic example for this type of NP is provided by a
model proposed by Banks and Zaks~\cite{Banks:1981nn}. 
In the following we will
denote the UV phase of the NP as the Banks-Zaks
($\bz$) phase, and the IR phase as the unparticle ($\ucal $) phase.

At energies below $\mc $ all mediator effects are virtual
and generate effective interactions between the new
physics  and the Standard Model (SM) sectors. The
dominant interactions are assumed to be of the form
\beq
 \lcal(UV) = c_\bz \mc^{-k} \ocal_\ssm \ocal_\bz 
\eeq
where $ \ocal_\ssm $ is a local, gauge-invariant  operator 
constructed out of the SM fields and their derivatives and,
similarly, $ \ocal_\bz $ denotes a local operator constructed
of the NP fields in the $\bz$ phase.
In the infrared these type of effective interactions
suffer from strong renormalization effects leading to the
replacement  $  \ocal_\bz \to \lu^{ \dbz - \du} \ocal_\ucal $,
where $ \dbz $ and $\du $ denote the scaling dimensions of the
operators $ \ocal_\bz$ and $ \ocal_\ucal $ respectively. 
We then have
\beq
\lcal(IR) = c_\ucal \mc^{-k} \ocal_\ssm 
\left( \lu^{ \dbz - \du} \ocal_\ucal \right) 
\eeq

The construction of the operator $ \ocal_\ucal $
is in general a difficult task, and requires detailed 
knowledge of the NP theory. However, for most calculations
this is not needed since one is interested in 
operator correlators and these are very strongly constrained
by conformal invariance. If one considers the effects of a single
operator $ \ocal_\ucal $ then for the purposes of calculating
cross sections  and for the effects on standard cosmology
one only needs the following density of states,
\bea
\int d^4 x e^{i q x }
\vevof{ \left[ \ocal_\ucal(x) , \ocal_\ucal(0) \right]}
 &=&  A_{\du} \theta(q^2) (q^2)^{\du-2} 
\left[ \theta(q^0) + \inv{ e^{ \beta |q_0| } - 1 } \right]
\eea
where $
A_n = (4 \pi)^{3-2n}/[2 \Gamma(n) \Gamma(n-1)] $ and
$ \vevof{\cdots} $ denotes the thermal average at
temperature $ 1/\beta $, and 
where we assumed $ \ocal_\ucal $ has bosonic character. 
For collider applications one simply lets $ \beta \to \infty $.
Using this expression one can determine the effects of this type
of NP on various experimentally interesting processes in terms
of a few parameters ($k,~\du,~\mc$ and $ \lu $). In the following
we  will
study some of the effects of this type of NP on standard cosmology 

\section{Thermodynamics}

\begin{figure}[ht]
\vspace{-1.5in}
\centering
\includegraphics[bb=0 0 300 500,width=5cm]{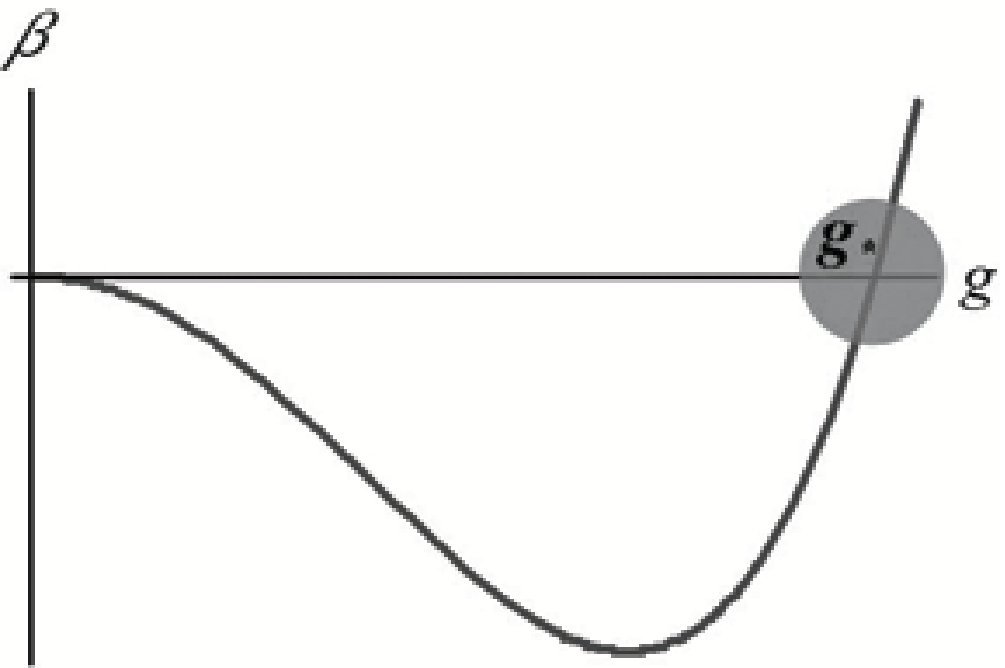} \qquad\qquad\qquad
\includegraphics[bb=0 0 300 500,width=5cm]{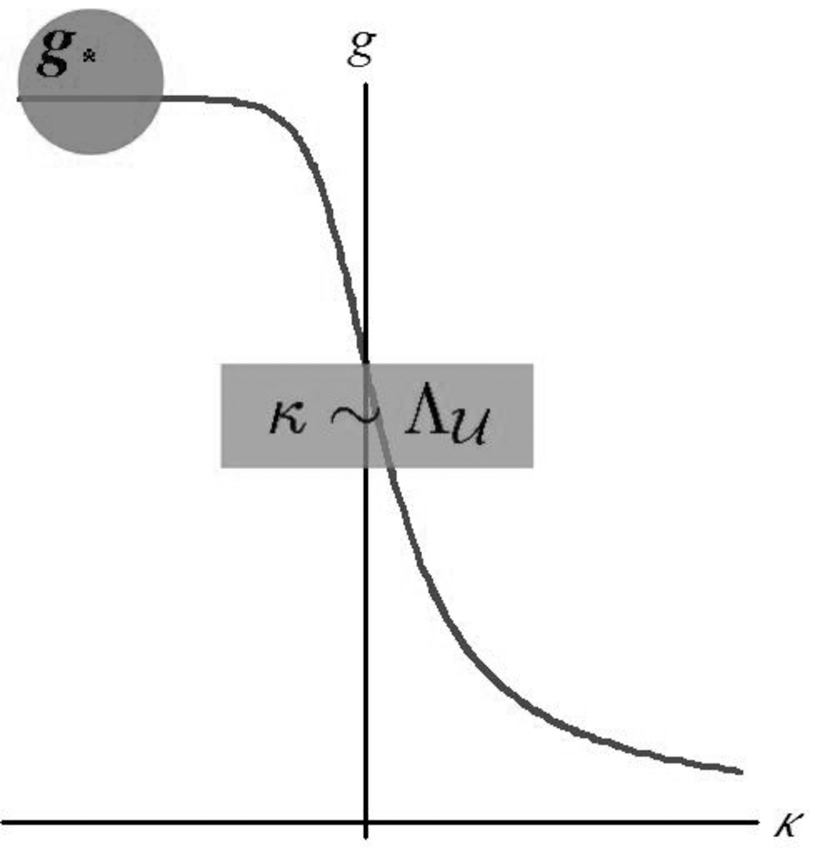}
\caption{Renormalization group behavior of unparticle
theories.}
\label{fig:f1}
\end{figure}

The conformal invariance requirement amounts to the assumption that
the NP beta function vanishes at $ g=g_*\neq 0$. 
In this case the $ \beta $ function and $g$  
will behave qualitatively as in Fig. \ref{fig:f1}.

For such theories the trace of the energy-momentum 
tensor~\cite{Collins:1976yq} obeys
\bea
\vevof{\theta_\mu^\mu} 
= \ru - 3 \Pu 
= (\beta/{2\lm}) \vevof{N \left[F^{\mu\nu}_a F_{a \; \mu\nu} \right] }
= (\beta/{2\lm}) b T^{4+\delta }
\eea
($N$ denotes normal ordering). It follows that $\langle\theta_\mu^\mu \rangle $
will vanish in the IR (since it is $\propto\beta$) and 
we expect $ \ru \simeq 3 \Pu $ at low temperatures. 
The leading corrections are produced by 
$ \vevof{N \left[F^{\mu\nu}_a F_{a \; \mu\nu} \right]} $, as indicated above,
with $ \delta $ the anomalous dimension of this operator. 

Using then standard thermodynamic relations we find
\beq
\ru     =   \sigma T^4   + A(1+3/\qwe) T^{4+\qwe} \,;\quad
\Pu     =   \sigma T^4/3 + (A/\qwe)    T^{4+\qwe}
\eeq
valid  in the IR regime. In the UV limit, we will
also have $ \rho_\snp \propto T^4 $ (up to logarithmic corrections)
since the theory is asymptotically free. Then
\beq
\rho = \frac3{\pi^2} \gnp T^4; \qquad \gnp = \left\{
\begin{array}{ll}
\gbz  & \bz~{\rm phase} \cr 
\gu  & \ucal~{\rm phase} 
\end{array} \right.
\eeq
where $ \gnp $ will be referred to as the number
of relativistic degrees of freedom (RDF).
For $ \bz $ models we have $ \gbz \sim 100 $,
both for the original example as well as others
that have been studied using lattice Monte Carlo 
methods~\cite{Svetitsky:2009pz}.
In the IR regime the RDF can be obtained using the
AdS/CFT correspondence~\cite{Gubser:1999vj} which gives
\beq
\gu  = (\pi^5/8) (L \mpl )^2 \gesim 100
\eeq
since $L$, the AdS radius is $ > 2/\mpl $.
It follows that  for all available models $ \gnp \gesim 100 $,
and in the IR we expect this number to be even larger.

\section{SM-NP interactions; equilibrium, freeze-out and thaw-in}

In order to understand the NP effects on cosmic evolution it is
important to determine when and if this sector was in equilibrium
with the SM sector. The standard approach for investigating this
makes heavy use of the Boltzmann equation (BE)~\cite{Kolb}, for our purposes
the idea is to {\it(i)} calculate $ \dot\rho_\ssm$ and $\dot\rho_\snp$ 
in terms of  $ \vartheta = \tnp - \tsm $ using the BE and then
{\it(ii)} use $ \rho \propto T^4 $ to obtain an evolution
equation of the form
\beq 
\dot\vartheta + 4 H \vartheta = - \Gamma \vartheta 
\eeq
together with an explicit and calculable expression for the reaction rate
$ \Gamma $.  $H$ denotes the Hubble parameter,
$ H^2 = [(8 \pi/(3 \mpl^2)] ( \rsm + \rnp ) $, where we assumed
a flat universe with zero cosmological constant. We will assume
that the SM-NP interactions are of the form
\beq
\lcal_{\rm int} = \epsilon \ocal_\ssm \ocal_\snp
\eeq

\begin{figure}[ht]
\vspace{-1.5in}
\centering
\includegraphics[bb=200 0 500 400,width=6cm]{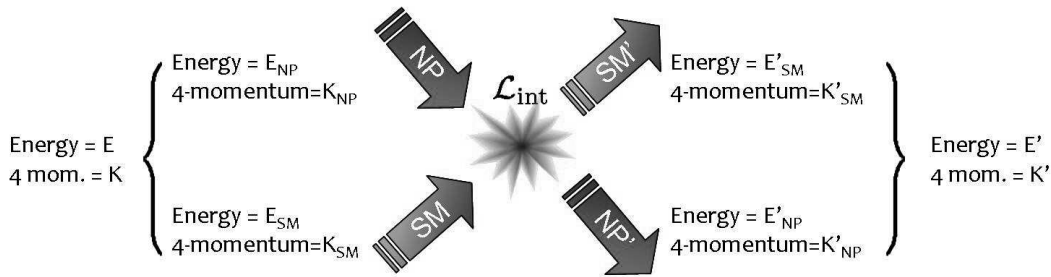} 
\caption{Definition of momenta involved in the BE calculation of $ \Gamma $.}
\label{fig:f2}
\end{figure}

The BE approach~\cite{Kolb} 
requires at intermediate steps the introduction of
the unparticle distribution function, which might pose conceptual 
problems. This can be avoided by using instead
the Kubo formalism~\cite{Kubo:1957mj}, though it lacks
the intuitive appeal of the BE. Both
formalisms yield identical expressions for $ \Gamma $
(the momenta in the BE expression are defined in 
Fig. \ref{fig:f2}):
\bea
\Gamma &=& \frac{\pi^2}{12 T^4} \left( \inv\gsm + \inv \gnp \right) \ical\,; \cr
\ical_{\rm Kubo}&=& \epsilon^2 
\Re\int_0^\beta ds \int_0^\infty dt \int d^3\xx\;
\vevof{\ocal_\ssm(-i s , \xx) \dot\ocal_\ssm(t,{\bf0})}
\vevof{\ocal_\snp(-i s , \xx) \dot\ocal_\snp(t,{\bf0})} \cr
\ical_{\rm BE} &=& \half  \sum
\int d\Phi_\snp d\Phi_\ssm \beta (E_\ssm - E_\ssm')^2 
e^{ - \beta E} \left| \mcal\right|^2 (2\pi)^4 \delta (K - K') 
\eea
and one can show $ \ical_{\rm BE}=\ical_{\rm Kubo}$. 

For the  arguments below the detailed structure of $ \Gamma $
will not be needed and the following order of magnitude estimate 
suffices:
\bea
\Gamma &\simeq& \frac{\epsilon^2 
\lambda (\gsm + \gnp)}{(4\pi)^{n_\ssm + n_\snp-1}}
 T^{2d_\ssm + 2d_\snp - 7} \qquad
\lambda =  \frac{\gsm'}\gsm \frac{\gnp'}\gnp
\eea
where $n_\ssm,~n_\snp $ are the number of  SM  and NP fields in 
$ \lcal_{\rm int} $, and $ \gsm',~ \gnp' $ are the RDF involved
in the interaction (in the unparticle phase we take 
$n_\snp = 2\du-2,~ \gnp' = \du$). It then follows that
\beq
\Gamma/ H \propto T^{2 d_\ssm + 2 d_\snp -9}
\eeq
where the SM and NP sectors will be (de)coupled as long as $ \Gamma > H $
($ \Gamma < H $), the transition temperature
$ \tf $ is determined by the condition $ \Gamma = H $. Note that 
when $ d_\ssm + d_\snp < 4.5 $, the sectors are coupled
for $ T$ {\em below} $ \tf $, what we call a ``thaw-in'' scenario;
in the complementary case $ d_\ssm + d_\snp > 4.5 $ decoupling
occurs for $ T < \tf $ (standard freeze-out scenario).

\begin{figure}[ht]
\vspace{2in}
\centering
\includegraphics[bb=100 0 300 200,width=5cm]{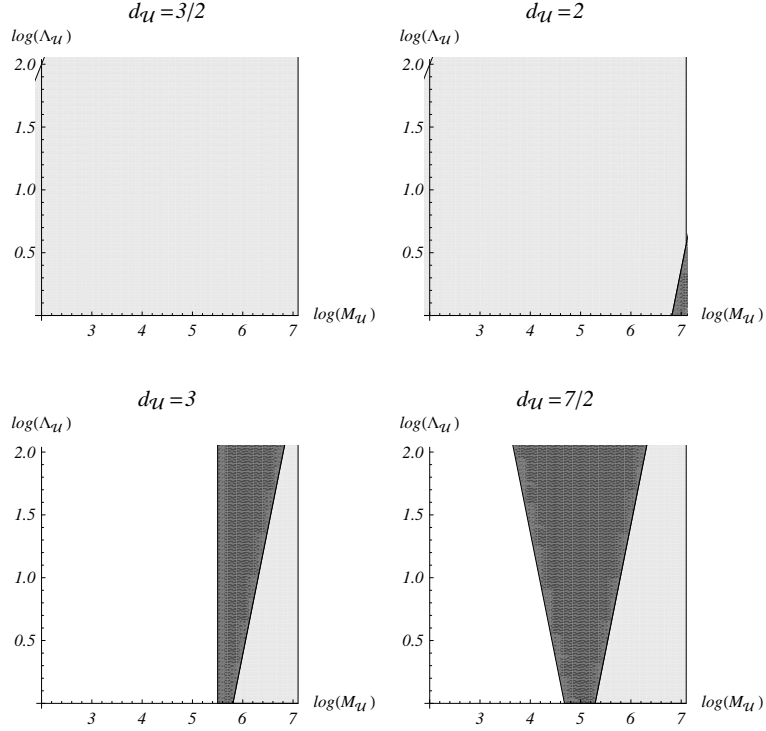}

\caption{Regions in the $\lu - \mc$ plane corresponding to various
freeze-out and thaw-in scenarios for $\du=3/2,~2,~3,~7/2$.
Dark grey:  SM-NP decoupling in the 
unparticle phase only; light gray: no SM-NP decoupling; in the white
regions $\tfu < v$ ($\lu, ~\mc$ are in TeV units).
We assumed $\gsm=\gbz=\gu=100$, $\gsm'=4$, $\gbz'=50$ and $\gu'=\du$.
$\bz$ phase: $n_{SM}=n_{NP}=2$, $\dsm=2$ and $d_{NP}=3$;
$ \ucal $ phase:  $n_{SM}=2$, $n_{NP}=2(\du-1)$, $\dsm=2$ and $d_{NP}=\du$.
}
\label{fig:f3}
\end{figure}

For the calculation we will use
\beq
\lcal_{\rm int} = \left\{ \begin{array}{ll}
 (\phi^\dagger \phi) \sum_Q \bar Q Q /\mc & \bz \cr 
 (\phi^\dagger \phi) \ocal_\ucal 
\left( \lu^{ 3- \du}/\mc \right) & \ucal
\end{array} \right.
\eeq
where $Q$ denotes a quark in the $ \bz $ phase of the NP sector and
$ \ocal_\ucal $ is the unparticle operator
corresponding to $ \sum_Q \bar Q Q $. For consistency we require
$ \mc > \tf > \lu $ in the $\bz$ sector, and $
\lu > \tf > v $ in the $\ucal$ sector (the lower value of $v$ 
is needed because below this scale the SM operator $ \phi^\dagger
\phi$ is no longer relevant).
With these preliminaries one can determine the regions in the
$ \mc - \lu $ for which the SM and NP are coupled in each
of the NP phases, the results are presented in Fig. \ref{fig:f3}

\section{Unparticle effects in Big Bang Nucleosynthesis (BBN)}

Let's first consider the case where SM and NP 
were in equilibrium down to a temperature $ T_f > v$, and decoupled
thereafter.
The relationship between the NP and SM temperatures are
thereafter determined by entropy conservation~\cite{Kolb}, specifically,
\bea
(T_f R_f)^3 g_{NP}^\star(T_f) &=& (T_{NP} R)^3 g_{NP}^\star(T_{NP}) \cr
(T_f R_f)^3 g_{SM}^\star(T_f) &=& (T_\gamma R)^3 g_{SM}^\star(T_\gamma)
\label{entcon}
\eea
where 
$\gnp^\star$ and $\gsm^\star$ stand for
the NP and SM effective numbers of RDF conventionally~\cite{Kolb}
adopted for the entropy density,  
$R_f,~(R) $ denote the scale factor at $T=T_f~(T_\gamma)$,
and $T_\gamma,~T_{NP}$ denote the SM and NP
temperatures at the BBN epoch.

Using this we find
\beq
T_{NP}=T_\gamma\left[\frac{g_\gamma}{g_\gamma+g_e} 
\frac{ g_\gamma+g_e+g_\nu}{g_{SM}(v)}\right]^{1/3}
\eeq
where $ g_\gamma = 2,\,
g_e = (7/8)\times 4 ,\,
g_\nu = (7/8) \times 3 \times 2 ,\,
g_\ssm(v) = 106.75 $.
The NP contribution to the total energy density
can be expressed in terms of an additional number of sterile
neutrinos $ \Delta N_\nu $ defined by the expression
\beq
\rnp =\frac{3}{\pi^2} \gir \tnp^4 
\equiv \frac{3}{\pi^2} \frac74 \left(\frac{4}{11}\right)^{4/3} 
\Delta N_\nu \; T_\gamma^4
\eeq
Then,
\beq
\gir = \frac74 \left[ 
\frac{g_{SM}(v)}{g_\gamma + g_e + g_\nu} \right]^{4/3} \Delta N_\nu
\eeq
Current data~\cite{Miele:2008qt} provides the limits 
$ \Delta N_\nu = 0 \pm 0.3 \pm 0.3 $, leading to
\beq
\gir < 20
\eeq
at the 95\% CL. 
In the extreme case where the SM and NP remain in equilibrium at the
BBN epoch this bound is considerably strengthened: $ \gir < 0.3 $.

Viable unparticle models should exhibit conformal invariance 
with a small number of RDF in the IR. We are unaware of any model 
with these characteristics. In fact the AdS/CFT correspondence
suggests this constraint is very strongly violated.

\section{Comments}
We have shown that strongly coupled NP can 
lead to $ \Gamma/H \sim T^n$ with $n$ positive or negative,
and that this results in a variety o freeze-out and thaw-in 
scenarios.
Current BBN data generates strong constraints on the
properties of the NP. Even for ``normal'' decoupling scenarios 
($n>0$) the BBN constraint is significant leading to a bound
$ \gir < 20 $ to be compared to $ \gir > 100  $ for the
available models.

Unparticle models also suffer from potential theoretical problems: 
the coupling to the SM necessarily breaks conformal invariance,
but the scale at which this occurs lies
below the BBN temperature for a range of $ \du $, so our
conclusions still apply in this case. 

\bce Acknowledgments \ece

We thank the Organizers of  
the XXXIII International Conference of Theoretical Physics
for their worm hospitality during the meeting.
This work is supported in part by the Ministry of Science and Higher
Education (Poland) as research project N~N202~006334 (2008-11) 
and by the U.S. Department of Energy
grant No.~DE-FG03-94ER40837.  B.G. acknowledges support of the
European Community within the Marie Curie Research \& Training
Networks: "HEPTOOLS" (MRTN-CT-2006-035505), and "UniverseNet"
(MRTN-CT-2006-035863), and through the Marie Curie Host Fellowships
for the Transfer of Knowledge Project MTKD-CT-2005-029466.

\end{document}